\documentclass[prl,twocolumn,showpacs,aps,amssymb,superscriptaddress]{revtex4-1}
\usepackage{graphicx}
\usepackage{epsfig}
\usepackage{color}
\usepackage{ulem}
\newcommand{\nn}[1]{{\langle #1 \rangle}}

\newcommand{\hc}{\mathrm{h.c.}}

\begin{document}

\title{Generic First-Order vs. Continuous Quantum Nucleation of Supersolidity}
\author{Lars Bonnes}
\affiliation{Institut f\"ur Theoretische Physik III, Universit\"at Stuttgart, Pfaffenwaldring 57, 70550 Stuttgart, Germany}
\author{Stefan Wessel}
\affiliation{Institut f\"ur Theoretische Physik III, Universit\"at Stuttgart, Pfaffenwaldring 57, 70550 Stuttgart, Germany}
\affiliation{Institute for Theoretical Solid State Physics, RWTH Aachen University, Otto-Blumenthal-Str. 26, 52056 Aachen}

\begin{abstract}
We analyze the nucleation of supersolid order out of the superfluid ground state of bosons on the triangular lattice. 
While the stability of  supersolidity against phase separation in this system is by now well established for nearest-neighbor and long-range dipolar
interactions, relevant for
two-dimensional arrays of ultra-cold polar molecules, here we address directly
the nature of the superfluid-to-supersolid  transition. 
Based on symmetry arguments and quantum Monte Carlo simulations,  we conclude that this quantum phase transition is driven first-order beyond the line of particle-hole symmetry.
Along this line, the transition is continuous and its scaling behavior  consistent with the three-dimensional (3D) XY universality class. We relate this finding to a 3D $\mathbb{Z}_6$ clock model description of the enlarged 
symmetry of the solid order parameter field. In the generic case however, the symmetry reduces to that of a 3D $\mathbb{Z}_3$ clock model, which reflects the first-order nature of the generic superfluid-to-supersolid quantum phase transition on the triangular lattice. 

\end{abstract}
\maketitle
Polar molecules~\cite{lahaye09} are considered
promising candidates for the realization of novel quantum states of matter. They feature inherently long-ranged dipolar interactions, and state dressing by static electric and microwave fields allows to tune the inter-particle potential over a wide range and even into a regime that is dominated by three-body interactions~\cite{micheli06,buechler07}. 
One  feasible scenario is the realization of a supersolid phase upon loading bosonic polar molecules on a triangular optical lattice~\cite{pollet10}. 
In fact, ensembles of hetero-nuclear molecules, such as KRb, RbCs, or LiCs
in the rovibronic ground-state have been produced~\cite{sage05,ni08,deiglmayr08,ospelkaus10,aikawa10}, and recently the superfluid to Mott insulator transition of ${}^ {87}$Rb atoms has been demonstrated  on a triangular optical lattice~\cite{becker10}. 
Already for nearest-neighbor repulsive interactions, geometric frustration on the triangular lattice gives
rise to an extended supersolid regime at sufficiently low kinetic energy~\cite{murthy97,wessel05a,heidarian05,melko05,boninsegni05,heidarian10}. 
This supersolid phase was found to be stable under extended dipolar interactions~\cite{pollet10} and for strong three-body repulsions that emerge between dressed polar molecules~\cite{bonnes11a}.

While such setups thus appear promising for an experimental realization of a supersolid state of matter, the nature of the transition from the superfluid to the supersolid regime, at which  solid order  nucleates, has not been systematically addressed thus far in previous numerical studies. Initial works concluded  the transition at or close to half-filling to be continuous, without however  specifying the universality class of the quantum phase transition~\cite{wessel05a,heidarian05,hassan07}. On the other hand, a recent cluster mean-field study concluded that the quantum phase transition is driven first-order beyond  half-filling, and exhibits anomalous hysteresis behavior~\cite{yamamoto11}. This calls for a careful re-examination of the quantum phase transition towards supersolidity on the triangular lattice.

\begin{figure}[t]
\begin{center}
\includegraphics[width=8.5cm]{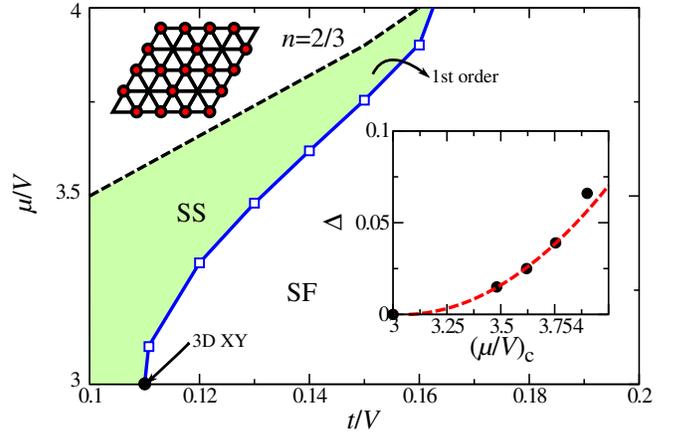}
\caption{(Color online) Ground state phase diagram for hard-core bosons with nearest neighbor repulsion on the triangular lattice near the superfluid-to-supersolid transition region. 
The inset shows 
the jump $\Delta$ of the  order parameter corresponding to the solid order shown in the upper inset  along the transition line as a function of the critical ratio $(\mu/V)_c$.
The dashed line in the inset is a quadratic fit for $\Delta<0.05$ to the deviation of $(\mu/V)_c$ from $3$ (half-filling).
Errors are below the symbol size.}
\end{center}
\end{figure}

Here, we employ large-scale quantum Monte Carlo (QMC) simulations to assess the above mentioned scenario of a first-order superfluid-to-supersolid transition. We find that  the transition is indeed first order beyond the particle-hole symmetric line of half-filling, while at  half-filling the transition is continuous, and consistent with the 3D XY universality class. We relate our numerical findings to the peculiar properties of supersolidity on the triangular lattice, the corresponding discrete ground state degeneracies and the symmetry of the effective order parameter field,  to describe the transition in terms of a $q$-state, $\mathbb{Z}_q$, clock model with $q=3$ (6) for the generic (particle-hole symmetric) case, respectively. 

In the following, we consider the basic model of hard-core bosons on the triangular lattice with nearest neighbor repulsions, described by the Bose-Hubbard like Hamiltonian
\begin{equation}
 H= -t \sum_\nn{ij} ({ b_i^\dagger b_j + \hc})
+ V \sum_\nn{ij} n_i n_j
- \mu \sum_i n_i.
\end{equation}
Here, $b_i$ ($b_i^\dagger$) annihilates (creates) a particle at lattice site $i$ and $n_i=b_i^\dagger b_i$ denotes the local density operator. 
The tunneling matrix element $t$ connects nearest-neighbor sites $\nn{ij}$, and $V$ denotes the repulsion between adjacent sites. Furthermore, $\mu$ denotes the chemical potential controlling the filling $n$ of the lattice. 
Before presenting our numerical results on the superfluid-to-supersolid transition, that are summarized in Fig.~1, we perform a symmetry analysis of the solid order across the superfluid-to-supersolid transition. 
The conclusions from these considerations are then shown to be consistent with the results from quantum Monte Carlo simulations.

The phase diagram of the model in Eq.~(1) has been subject to various analytical~\cite{murthy97,burkov05,hassan07} and numerical studies \cite{wessel05a,heidarian05,melko05,boninsegni05}. For small values of $t/V$,  two incompressible solids form, with $\sqrt{3} \times \sqrt{3}$ diagonal long-range order (DLRO) of filling $1/3$ and $2/3$ respectively, related by particle-hole symmetry around $\mu/V=3$. 
For large values of $t/V$, a uniform superfluid (SF) persists,  where the system exhibits off-diagonal long-range order (ODLRO) and  a finite superfluid density $\rho_s$.
Supersolid (SS) phases with both DLRO and ODLRO appear between the two solid lobes and prevail down to $t/V=0$ at $\mu/V=3$. These supersolid phases are characterized by local density fluctuations $(\epsilon_1, \epsilon_2, \epsilon_3)=(\mp2 \epsilon, \pm \epsilon', \pm \epsilon')$ around a triangular plaquette, where $\epsilon_i=\langle n_i - \langle n \rangle \rangle$ . The different signs relate to supersolids of filling $n<1/2$ (SS-A) and $n>1/2$ (SS-B), respectively.
At half filling ($\mu/V=3$), both supersolid phases coexist at a first-order SS-SS transition where both $S/N$ and $\rho_s$ are continuous, whereas the pattern of the density fluctuations around a plaquette changes from SS-A to SS-B~\cite{boninsegni05}. 

In terms of the phase $\theta$ of the complex order parameter 

$
\epsilon e^{i\theta}=\epsilon_1+\epsilon_2 e^{-i2\pi/3}+\epsilon_3 e^{i2\pi/3},
$

the two supersolids are distinct in that $\nn{\cos (3\theta)} < 0$ for SS-A and $\nn{\cos (3\theta)}>0$ for SS-B~\cite{heidarian05}.
Furthermore, this quantity respects the three-fold rotational symmetry of the lattice ($\theta \rightarrow \theta+2\pi/3$).
We thus expect that
the nucleation of solid order at the generic (i.e. $\mu/V\neq 3$) SF-SS quantum phase transition relates to that of a 3D three-state, $\mathbb{Z}_3$, clock model. 
The $\mathbb{Z}_3$ clock model equals a three-state Potts model~\cite{hove03}, which in 3D
undergoes a first-order ordering transition~\cite{wu82}. Thus the generic SF-SS quantum phase transition is expected to be first-order.

At $\mu/V=3$ however, where the SS-A and SS-B coexist, one finds  $\nn{\cos (3\theta)} = 0$, while $\nn{\cos (6\theta)} > 0$. The symmetry is thus enhanced to a six-fold rotation of the order parameter by $\pi/3$.
This symmetry enhancement has consequences on the nature of the SF-SS phase transition, since the symmetry breaking now relates to that of a six-state, $\mathbb{Z}_6$, clock model.
The 3D $\mathbb{Z}_6$ clock model has a continuous transition within the 3D XY universality class~\cite{blankschtein84,oshikawa00,hove03,lou07}. Hence, we conclude that the SF-SS quantum phase transition at  $\mu/V=3$ is continuous and within the 3D XY universality class. Thus far, in our analysis we neglected the coupling between the solid order and superfluidity; however, as argued in Ref.~\cite{frey97}, the 3D XY fixed-point is stable in the presence of such interactions.
From these results, we expect that the line of first-order SF-SS transition has a critical endpoint at $\mu/V=3$. In this scenario, the continuous transition at the particle-hole symmetric point of half-filling is not driven first-order via fluctuations~\cite{coleman73}, but turns discontinuous due to a reduced order parameter symmetry beyond the line of half-filling. 

\begin{figure}[t]
\begin{center}
\includegraphics[width=8.5cm]{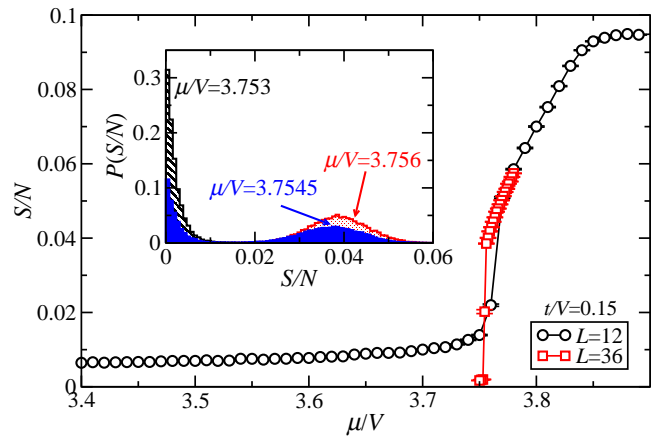}
\caption{(Color online) Structure factor $S/N$ for $t/V=0.15$ obtained at $\beta=100$ for $L=12$ and $36$. Inset: histograms of $S/N$  across the SF-SS transition for $L=36$.}
\end{center}
\end{figure}

We next employ unbiased QMC simulations based on a directed-loop algorithm in the stochastic series expansion representation~\cite{sandvik99b,syljuasen02,alet05} to assess the SF-SS transition. The interactions are decoupled into three-site triangular plaquettes, which  allows for efficient updates. Our simulations are performed on systems with linear system size $L$ up to $54$ (the number of sites $N=L^2$), employing periodic boundary conditions in both lattice directions. Ground-state properties are obtained by tuning the temperature $T$ sufficiently low, as detailed below. The superfluid density $\rho_s$ is obtained from measuring the winding number fluctuations~\cite{pollock87}. DLRO is detecting by measuring the corresponding structure factor 
$
S=1/N \sum_{ij} e^{i \mathbf{Q} (\mathbf{r}_{i}-\mathbf{r}_{j})} \langle n_i\: n_j \rangle
$
at wave vector $\mathbf{Q} \equiv (2 \pi / 3, 0)$, where the position of lattice site $i$ is denoted $\mathbf{r}_{i}$.
Since the problem is particle-hole symmetric, we need to consider only the regime $\mu/V \ge 3$.

%{\it Results}.--
Distinguishing  a (weakly) first-order from a continuous (quantum) phase transition can be challenging due to restrictions in the accessible lattice sizes with respect to the relevant correlation length.
We thus start our analysis in a regime well away from half-filling, i.e. in the vicinity of the solid phase. Fig.~2 shows the structure factor $S/N$ for $t/V=0.15$, upon driving the chemical potential 
through the SF-SS quantum phase transition. A pronounced jump in $(S/N)$ and $\rho_s$ (not shown) develops as one increases the system size, characteristic of a first-order transition.
The inset in Fig.~2 shows histograms recorded in the close vicinity of the  transition point, displaying a robust two-peak structure at $(\mu/V)_c\approx 3.7545$.
This indicates that the ordered and disordered phase coexist at this point.
The jump in the order parameter of about $\Delta\approx 0.04$ is estimated from the position of the peak at finite $S/N$, cf. also the inset of Fig.~1.

\begin{figure}[t]
\begin{center}
\includegraphics[width=8cm]{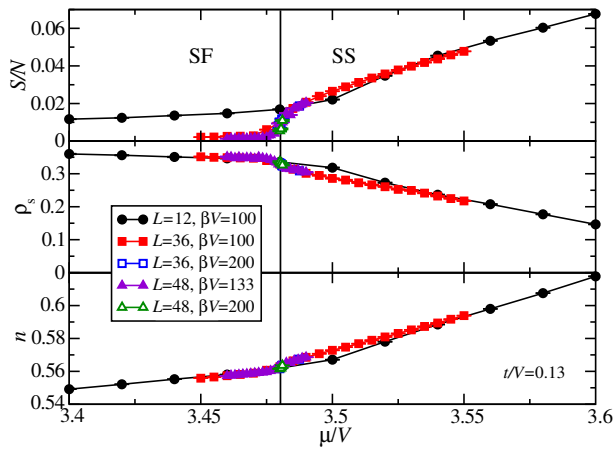}
\caption{(Color online) Filling $n$, superfluid density $\rho_s$ and structure factor $S/N$ for $t/V=0.13$  for various system sizes as functions of $\mu/V$ across the 
SF-SS transition.}
\end{center}
\end{figure}

Moving in parameter space towards $\mu/V=3$, the DLRO (and also $\rho_S$) weakens, and the jump $\Delta$ in the order parameter decreases, as seen from the inset of Fig.~1.
In particular, 
Fig.~3 shows the filling $n$, the superfluid density $\rho_s$ and the order parameter $S/N$ as functions of $\mu/V$ for $t/V=0.13$, in the vicinity of the SF-SS quantum phase transition. 
The data appears to vary smoothly across the transition, in pronounced contrast to  the results shown in Fig.~2 for $t/V=0.15$.
The histograms of $S/N$ Fig.~4 for sufficiently large systems, however, reveal  the emergence of a two-peak structure at $(\mu/V)_c\approx 3.48075$, which remains stable in the thermodynamic limit, 
as shown in the inset of Fig.~4. 
We estimate the jump in the order parameter at the transition point for  $t/V=0.13$ as $\Delta\approx 0.0125$ form Fig.~4. 
We also performed simulations at fixed $t/V=0.12$ and $\mu/V=3.1$, but we were not able to observe pronounced features in the $S/N$ histograms apart from  only a slight broadening of the peak in $S/N$, at least within the accessible systems sizes up to $L=60$.
%Hence, we cannot decide on the nature of the SF-SS transition for $t/V\lesssim 0.12$. 
In fact, based on an approximate quadratic scaling 
$
\Delta\propto \left( \left(\mu/{V}\right)_c - 3\right)^2
$
observed in the numerical data near half-filling (cf. the fit in the inset of Fig.~1), we conclude that in this regime
the first order transition has become extremely weak, with an estimated $\Delta \lesssim 0.002$ at $t/V=0.12$, and is thus essentially resolution limited on the accessible system sizes. 
\begin{figure}[t]
\begin{center}
\includegraphics[width=8.5cm]{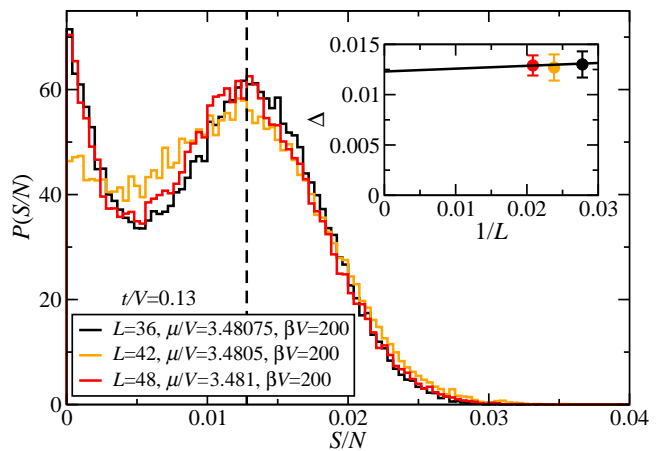}
\caption{(Color online) Histograms of $S/N$ obtained at $t/V=0.13$ for different system sizes. Inset:  Extrapolation of the peak position $\Delta$ to the thermodynamic limit.}
\end{center}
\end{figure}
We thus obtained clear evidence for a first-order SF-SS transition over a wide range of the phase diagram away from half-filling, with a decreasing discontinuity  upon approaching the half-filled limit. 
Next, we focus on the SF-SS transition directly at $\mu/V=3$, where based on  symmetry considerations the quantum phase transition is 
expected to be continuous. Consistent with previous simulations~\cite{wessel05a,heidarian05,melko05}, 
we obtain no indication for a first-order transition at $\mu/V=3$ within the range of considered system sizes (up to $L=54$).
In order to assess, if the transition indeed is of 3D XY type, 
we employ finite-size scaling analysis. 
The finite-size data of $S/N$ for different system sizes in the vicinity of the SF-SS quantum phase transition is shown in the left inset of Fig.~5. Related to  the dynamical critical exponent $z=1$~\cite{frey97}, 
we performed the simulations fixing $TL=0.1V$.  
We extract the critical exponents using a conventional finite-size scaling ansatz for the 
structure factor $S/N$ and the susceptibility $\chi=(1/N)\int d\tau \langle\sum_{i,j} e^{i{\mathbf Q}({\mathbf r_i}-{\mathbf r_j})}   n_i(\tau) n_j(0)\rangle$ at $\mathbf{Q} \equiv (2 \pi / 3, 0)$, 
where $n_i(\tau)=e^{iH\tau} n_i e^{-iH\tau}$.
The finite system data for $S/N$ near a quantum critical point should follow the scaling relation 
$
S/N= L^{-2\beta/\nu} g\left(\frac{\tau-\tau_c}{\tau_c} L^{1/\nu}\right).
$
Similarly, we extract $\gamma$ and $\nu$ from the scaling of $\chi$. 
Here, $\tau=t/V$ and $\tau_c$ denotes the position of the quantum critical point.
From our analysis, we locate the critical point at 
$(t/V)_c=0.1108(2)$
and obtain the critical exponents 
$\nu=0.67(2)$, $\beta=0.32(2)$ from the scaling of the structure factor, 
and $(t/V)_c=0.1105(3)$ and $\nu=0.68(4)$, $\gamma=-1.3(2)$ from the 
data collapse of the susceptibility. The error bars were obtained using standard bootstrapping, cf. the right inset of Fig.~5 for typical distributions.
These results are consistent with recent high-precision values 
$\beta=0.3486(1)$, $\gamma=-1.3178(2)$ and $\nu=0.6717(1)$
of the 3D XY universality class~\cite{campostrini06}.
The raw data and the data collapse using the reference values of the  critical exponents are shown in Fig.~5.

\begin{figure}[t]
\begin{center}
\includegraphics[width=8.5cm]{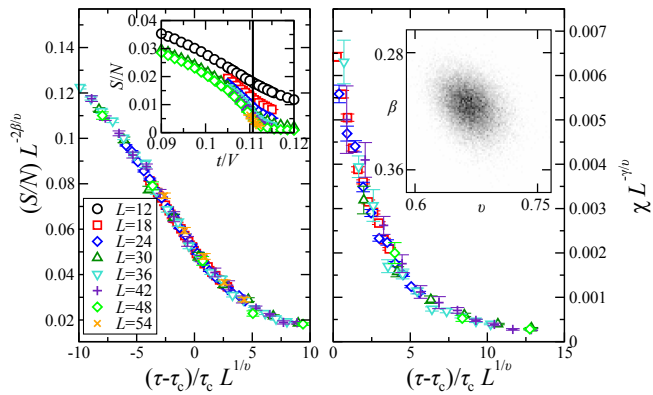}
\caption{(Color online) Data collapse of the structure factor $S/N$ (left panel) and the susceptibility $\chi$ (right panel) at $\mu/V=3$ for system sizes from $L=12$ to $54$ at $LT=0.1V$. 
Left inset: finite-size data of $S/N$. Right inset: bootstrapping histogram for the critical exponents $\beta$ and $\nu$ from $S/N$ data.
}
\end{center}
\end{figure}

In conclusion, 
we analyzed  the nucleation of supersolidity of hard-core bosons on the triangular lattice;
based on a symmetry analysis and quantum Monte Carlo simulations, we found that this transition is first-order beyond the line of half-filling, whereas at half-filling, the transition is continuous and consistent with the 3D XY universality class.
It will be important to revisit the quantum phase transitions between superfluid and supersolid phases also on other lattice geometries:
For example, in case of a spin-1/2 bilayer system, the superfluid-to-supersolid transition was reported from numerical simulations to belong to the 3D Ising universality class~\cite{laflorencie07}, in contrast to the field-theoretical expectation for a checkerboard supersolid state~\cite{frey97,schmidt08}. It will be particularly interesting to address the question, if the transition between the checkerboard supersolid state and the superfluid is in fact driven first-order by fluctuations for generic fillings, as suggested from an $\epsilon$-expansion\cite{frey97}, or if the exotic non-Bose-liquid fixed-point obtained from a fixed-dimensional renormalization group analysis~\cite{frey97} could be realized. 
Our results imply that on the triangular lattice supersolid domains proliferate upon driving a superfluid system across the transition to the supersolid region -- the need to distinguish such a phase coexistence from a mere mixture of (non-superfluid) solid and (non-solid) superfluid domains (like at a first-order superfluid-to-solid transition) could well challenge a robust experimental identification of the emerging supersolid state.

\begin{acknowledgements}
We thank H.-P. B\"uchler and X.-F. Zhang
for discussions, and
acknowledge the allocation of CPU time on the HLRS Stuttgart
and NIC J\"ulich supercomputers.
Support was also provided through
the Studienstiftung des Deutschen Volkes (LB)
and
the DFG within SFB/TRR 21 (SW).
\end{acknowledgements}


\begin{thebibliography}{10}

\bibitem{lahaye09}
T. Lahaye, {\it et al.} Rep. Prog. Phys. {\bf 72},  126401 (2009).

\bibitem{micheli06}
A. Micheli, G.~K. Brennen, and P. Zoller, Nature Physics {\bf 2},  341  (2006).

\bibitem{buechler07}
H.~P. B{\"u}chler, A. Micheli, and P. Zoller, Nature Physics {\bf 3},  726 (2007).

\bibitem{pollet10}
L. Pollet, J.~D. Picon, H.P. B{\"u}chler, and M. Troyer, Phys. Rev. Lett. {\bf 104},  125302  (2010).

\bibitem{sage05}
J.~M. Sage, S. Sainis, T. Bergeman, and D. DeMille, Phys. Rev. Lett. {\bf 94}, 203001  (2005).

\bibitem{ni08}
K.-K. Ni, {\it et al.}, Science {\bf 322},  231  (2008).

\bibitem{deiglmayr08}
J. Deiglmayr, A. Grochola, M. Repp, K. Mortlbauer, C. Gluck, J. Lange, O. Dulieu, R. Wester, and M. Weidem{\"u}ller, Phys. Rev. Lett. {\bf 101},  133004  (2008).

\bibitem{ospelkaus10}
S. Ospelkaus, {\it et al.} Phys. Rev. Lett. {\bf 104}, 030402 (2010).

\bibitem{aikawa10} 
K. Aikawa, {\it et al.}, Phys. Rev. Lett. {\bf 105}, 203001 (2010).

\bibitem{becker10}
C. Becker, {\it et al.}, New J. Phys. {\bf 12}, 065025 (2010).

\bibitem{murthy97}
G. Murthy, D. Arovas, and A. Auerbach, Phys. Rev. B {\bf 55},  3104  (1997).

\bibitem{wessel05a}
S. Wessel and M. Troyer, Phys. Rev. Lett. {\bf 95},  127205  (2005).

\bibitem{heidarian05}
D. Heidarian and K. Damle, Phys. Rev. Lett. {\bf 95},  127206  (2005).

\bibitem{melko05}
R.~G. Melko {\it et~al.}, Phys. Rev. Lett. {\bf 95},  127207  (2005).

\bibitem{boninsegni05}
M. Boninsegni and N. Prokof'ev, Phys. Rev. Lett. {\bf 95},  237204  (2005).

\bibitem{heidarian10}
D. Heidarian and A. Paramekanti, Phys. Rev. Lett. {\bf 104}, 015301 (2010).

\bibitem{bonnes11a}
L. Bonnes and S. Wessel, Phys. Rev. B. {\bf 83}, 134511 (2011).

\bibitem{hassan07}
S.~R. Hassan, L. {de Medici}, and A.-M.~S. Tremblay, Phys. Rev. B {\bf 76}, 144420  (2007).
  
\bibitem{yamamoto11}
D. Yamamoto, I. Danshita, and C. A. R. S{\'{a}} de Melo, arXiv:1102.1317v1.

\bibitem{burkov05}
A. Burkov and L. Balents, Phys. Rev. B {\bf 72},  134502  (2005).

\bibitem{hove03}
J. Hove and A. Sudb\o{}, Phys. Rev. E {\bf 68}, 046107 (2003).

\bibitem{wu82}
F.~Y. Wu, Rev. Mod. Phys. {\bf 54},  235  (1982).

\bibitem{blankschtein84}
D. Blankschtein {\it et~al.}, Phys. Rev. B {\bf 29}, 5250 (1984).

\bibitem{oshikawa00}
M. Oshikawa, Phys. Rev. B {\bf 61}, 3430 (2000).

\bibitem{lou07}
J. Lou, A.~W. Sandvik, and L. Balents, Phys. Rev. Lett. {\bf 99},  207203
  (2007).

\bibitem{frey97}
E. Frey and L. Balents, Phys. Rev. B {\bf 55},  1050  (1997).

\bibitem{coleman73}
S. Coleman and E. Weinberg, Phys. Rev. D {\bf 7},  1888  (1973);
B.~I. Halperin, T.~C. Lubensky, and S.-K. Ma, Phy. Rev. Lett. {\bf 32}, 292  (1974).


\bibitem{sandvik99b}
A.~W. Sandvik, Phys. Rev. B {\bf 59},  R14157  (1999).

\bibitem{syljuasen02}
O.~F. Sylju{\aa}sen and A.~W. Sandvik, Phys. Rev. E {\bf 66},  046701  (2002).

\bibitem{alet05}
F. Alet, S. Wessel, and M. Troyer, Phys. Rev. E {\bf 71},  036706  (2005).

\bibitem{pollock87}
E.~L. Pollock and D.~M. Ceperley, Phys. Rev. B. {\bf 36},  8343  (1987).

%\bibitem{beach05}
%K.~S.~D. Beach, L. Wang, and A.~W. Sandvik, Report cond-mat:0505194  (2005).

%\bibitem{wang06}
%L. Wang, K. Beach, and A. Sandvik, Phys. Rev. B {\bf 73},  014431  (2006).

%\bibitem{wenzel08}
%S. Wenzel, L. Bogacz, and W. Janke, Phys. Rev. Lett. {\bf 101},  127202 (2008).

%\bibitem{wenzel09}
%S. Wenzel and W. Janke, Phys. Rev. B {\bf 79},  014410  (2009).

\bibitem{campostrini06}
M. Campostrini, {\it et al.}, Phys. Rev. B {\bf 74}, 144506 (2006).

\bibitem{laflorencie07}
N. Laflorencie and F. Mila, Phys. Rev. Lett. {\bf 99},  027202  (2007).

\bibitem{schmidt08}
K. P. Schmidt,  {\it et al.}, Phys. Rev. Lett. {\bf 100}, 090401 (2008).

\end{thebibliography}
\end{document}